\documentclass[aps,pre,twocolumn,groupedaddress,showpacs]{revtex4}
\begin{document}

\title{Long-time dynamics of Rouse-Zimm polymers in dilute solutions with hydrodynamic memory}%

\author{V. Lisy$^{a)}$} \author{J. Tothova}
\affiliation{Institute of Physics, P.J. Safarik University,
Jesenna 5, 041 54 Kosice, Slovakia}

\author{A.V. Zatovsky}
\affiliation{Department of Theoretical Physics, I.I. Mechnikov
Odessa National University, 2, Dvoryanskaya Str., 65026 Odessa,
Ukraine}
\date{\today}%

\begin{abstract}
The dynamics of flexible polymers in dilute solutions is studied
taking into account the hydrodynamic memory, as a consequence of
fluid inertia. As distinct from the Rouse-Zimm (RZ) theory, the
Boussinesq friction force acts on the monomers (beads) instead of
the Stokes force, and the motion of the solvent is governed by the
nonstationary Navier-Stokes equations. The obtained generalized RZ
equation is solved approximately. It is shown that the time
correlation functions describing the polymer motion essentially
differ from those in the RZ model. The mean-square displacement
(MSD) of the polymer coil is at short times $\sim t^{2}$ (instead
of $\sim t$). At long times the MSD contains additional (to the
Einstein term) contributions, the leading of which is $\sim
\sqrt{t}$. The relaxation of the internal normal modes of the
polymer differs from the traditional exponential decay. It is
displayed in the long-time tails of their correlation functions,
the longest-lived being $\sim t^{-3/2}$ in the Rouse limit and
$t^{-5/2}$ in the Zimm case, when the hydrodynamic interaction is
strong. It is discussed that the found peculiarities, in
particular an effectively slower diffusion of the polymer coil,
should be observable in dynamic scattering experiments.

\end{abstract}

\pacs{36.20.Ey, 82.35.Lr, 83.80.Rs, 05.40.Jc}

\maketitle
\newpage
\section{Introduction}
Most of the theoretical investigations on the dynamic properties
of flexible macromolecules performed so far are based on the
Rouse-Zimm (RZ) model ~\cite{1, 2, 3, 4}. In this theory the
polymer molecule is modeled as a chain of beads under Brownian
motion. The bead participates in the interactions with solvent
molecules. The solvent contributes a frictional force against the
motion of a monomer and a random force to take into account the
random collisions exerted on monomers. In the Rouse model, the
solvent is considered nonmoving. Within the Zimm theory, the
motion of each monomer also affects other monomers, via the flow
which it induces in the ambient medium. It has been shown that the
Zimm model predicts the correct dynamical behavior for dilute
polymer solutions in $\theta$-conditions. The Rouse model is
applicable for good solvents, where the corrections due to the
hydrodynamic interactions and excluded volume effects  cancel each
other to a large extent, or in the situations when the surrounding
polymers screen out the hydrodynamic interactions ~\cite{3,4}.
Both models assume Gaussian equilibrium distribution of the beads,
which is in most cases the only description of a polymer that can
be manipulated analytically. The models hold for such polymer
properties, which involve length scales that are large compared to
monomer sizes. Although the RZ model is considered a universal
theory well describing the long-time behavior of the polymer
macromolecules, there is still a number of unresolved problems in
the understanding of the polymer dynamics in solution ~\cite{5,6}.
So, the $q^{3}$ dependence of the first cumulant in the
intermediate scattering vector regime is well confirmed for
synthetic polymers and DNA, however, the experimental values are
smaller than the theoretical predictions. Systematic deviations
from the theoretical behavior at large scattering vectors $q$ have
been found for various polymers using quasi-elastic neutron
scattering. At small $q$ the diffusion coefficient calculated from
the continuous RZ model also deviates from the experimental values
for both the natural and synthetic macromolecules. For a review of
these and other difficulties of the RZ theory we refer also to the
earlier work ~\cite{7}; it can be seen that the situation has
changed little during a decade.\

In this work we propose a generalization of the RZ theory that
could provide a better understanding of the dynamical behavior of
polymers in dilute solutions. The bead and fluid inertia during
the motion of the polymer in the solvent have been taken into
account. Also the hydrodynamic interaction is considered by
solving the nonstationary Navier-Stokes equations. A possible
importance of such a development (in connection with the Rouse
model) was already discussed in Ref. 8. As distinct from the
traditional approach to the polymer dynamics ~\cite{1, 2, 3, 4},
in our theory the resistance force on the moving bead will not be
assumed the Stokes one, which is proportional to the velocity of
the bead. We model this force by the Boussinesq-Basset history
force ~\cite{9, 10, 11} which, at a given time $t$, is determined
by the state of the bead motion in all the preceding moments of
time. We thus have a possibility to obtain solutions valid for
almost arbitrary $t$ (for incompressible fluids, however,
$t>>b/c$, where $b$ is the monomer size and $c$ the sound
velocity). The discussed generalization means that the effects of
hydrodynamic memory are taken into account. Such effects are
extensively studied in the physics of simple liquids and in the
theory of Brownian motion (see e.g. Ref. ~\cite{12}) with very
important consequences. In particular, the memory effects are
revealed in such interesting peculiarities like the famous
"long-time tails" of the molecular velocity autocorrelation
function (VAF), first discovered by means of computer experiments
~\cite{13,14}. The concept of the Brownian motion lies in the
basis of the RZ theory of polymer dynamics. Thus it is natural to
expect that the memory effects are important to polymers as well.
An attempt to show it was done in our recent paper ~\cite{15}
where the Rouse model has been considered taking into account the
hydrodynamic memory but with no hydrodynamic interaction. Here we
present a more general theory that, as limiting cases, includes
both the Rouse and Zimm models. It will be shown that the
inclusion of the hydrodynamic memory into the classical RZ model
leads to an essentially different behavior of the time correlation
functions describing the polymer motion in solution. It will be
demonstrated by the appearance of long-time tails of these
functions that reflect a strong persistence of the correlation
with the initial state of the polymer. The relaxation of the mean
square displacement of the whole polymer, as well as the decay of
the polymer internal modes, are slower than in the original model.
We also show how the tails are displayed in the long-time behavior
of the Van Hove function used in the interpretation of
quasi-elastic scattering of light and neutrons. It is discussed
that the found new features in the polymer dynamics should be
observable in the scattering experiments. Moreover, we believe
that the obtained results could contribute to the solution of some
of the existing problems between the theory and experiment.

\section{The dynamics of polymers with hydrodynamic memory}
Within the RZ model the motion of the $n$th polymer segment (the
bead) of a flexible polymer coil consisting of $N$ beads is
described by the equation
\begin{equation} \label{eq1}
M\frac{d^{2}\overrightarrow{x}_{n}(t)}{dt^{2}}=\overrightarrow{f}_{n}^{\text{fr}}
+\overrightarrow{f}_{n}^{\text{ch}}+\overrightarrow{f}_{n}.
\end{equation}
Here, $\overrightarrow{x}$ is the position vector of the bead, $M$
is its mass, $\overrightarrow{f}_{n}^{\text{ch}}$is the force from
the neighboring beads along the chain, $\overrightarrow{f}_{n}$ is
the random force due to the motion of the molecules of solvent,
and $\overrightarrow{f}_{n}^{\text{fr}}$ is the friction force on
the bead during its motion in the solvent. In the RZ model the
latter force was
\begin{equation}\label{eq2}
\overrightarrow{f}_{n}^{\text{fr}}=-\xi
\left[\frac{d\overrightarrow{x}_{n}}{dt}-\overrightarrow{v}\left(\overrightarrow{x}_{n}\right)\right],
\end{equation}
with $\overrightarrow{v}\left(\overrightarrow{x}_{n}\right)$ being
the velocity of the fluid in the place of the $n$the bead due to
the motion of other beads. The friction coefficient for a
spherical particle of radius $b$ is $\xi=6\pi\eta b$, where $\eta$
is the solvent viscosity. However, this expression holds only for
the steady-state flow. In the general case the resistance on the
body moving in a liquid depends on the whole history of the
motion, i.e. on velocities and accelerations in the preceding
moments of time. For incompressible fluids we use the Boussinesq
force ~\cite{9,16} instead of Eq. (2). This means that we replace,
in the Fourier transformation (FT) with respect to the time, the
friction coefficient $\xi$ with the frequency dependent quantity
\begin{equation}\label{eq3} \xi^{\omega}=\xi \left[1+\chi
b+\frac{1}{9}\left(\chi b \right)^{2}\right],
\end{equation}
where $\chi=\sqrt{-i\omega \rho/\eta}$, $(\text{Re} \chi >0)$ and
$\rho$ is the density of the solvent. The Boussinesq force follows
naturally from the usual hydrodynamics as the solution of
linearized Navier-Stokes equations ~\cite{17,18}. Additionally to
the Stokes force it contains terms  which, if the fluid density is
comparable to the density of the bead, cannot be neglected for the
nonstationary motion when they are of the same order as the
inertial term in Eq. (1). Equations (1-3) have to be solved
together with the hydrodynamic equations for the velocity of the
solvent,\begin{equation}\label{eq4} \rho \frac{\partial
\overrightarrow{v}}{\partial t}=-\nabla p+\eta \triangle
\overrightarrow{v}+\overrightarrow{\varphi},\\\
     \nabla\overrightarrow{v}=0.
\end{equation}
Here $p$ is the pressure. The quantity $\overrightarrow{\varphi}$
is an external force per unit
volume~\cite{4},\begin{equation}\label{eq5}
\overrightarrow{\varphi}\left(\overrightarrow{x}\right)=-\sum_{n}\overrightarrow{f}_{n}^{\text{fr}}
\left(\overrightarrow{x}_{n}\right)\delta\left(\overrightarrow{x}-\overrightarrow{x}_{n}\right).
\end{equation}
Equations (4) are solved using the FT in coordinates and time. The
solution can be, for any of the components $\alpha = x, y$, or $z$
written in the form
\begin{equation}\label{eq6}
v_{\alpha}^{\omega}\left(\overrightarrow{r}\right)=\int
d\overrightarrow{r}'\sum_{\beta}H_{\alpha\beta}^{\omega}\left(
\overrightarrow{r}-\overrightarrow{r}'\right)\varphi_{\beta}^{\omega}\left(\overrightarrow{r}'\right),
\end{equation}
with the FT of the Oseen tensor
\begin{equation}\label{eq7}
H_{\alpha\beta}^{\omega}\left(\overrightarrow{r}\right)=A\delta_{\alpha\beta}+Br_{\alpha}r_{\beta}r^{-2},
\end{equation}\begin{eqnarray}\label{eq8}
A=\left(8\pi\eta r\right)^{-1}
\{e^{-y}-y\left[\left(1-e^{-y}\right)y^{-1}\right]''\},\nonumber\\
B=\left(8\pi\eta r\right)^{-1}
\{e^{-y}+3y\left[\left(1-e^{-y}\right)y^{-1}\right]''\}.
\end{eqnarray}
Here the prime denotes the differentiation with respect to
$y=r\chi$. Substituting $\varphi_{\beta}^{\omega}$ from the FT of
Eq. (5) to $v_{\alpha}^{\omega}$ from (6), and the obtained result
into the FT of equation of motion (1), we get a generalization of
the RZ equation, which in the continuum approximation
reads\begin{eqnarray}\label{eq9} -i\omega
x_{n\alpha}^{\omega}=\frac{1}{\xi^{\omega}}\left[f_{\alpha}^{\text{ch,}
\omega}\left(n\right)
+f_{\alpha}^{\omega}\left(n\right)+M\omega^{2}x_{\alpha}^{\omega}\left(n\right)\right]
\end{eqnarray}\begin{eqnarray}
+\int_{0}^{N}dm H_{\alpha\beta
nm}^{\omega}\left[\frac{3k_{B}T}{a^{2}}\frac{\partial^{2}x_{\beta}^{\omega}}{\partial
m^{2}}+f_{\beta}^{\omega}\left(m\right)+M\omega^{2}x_{\beta}^{\omega}\left(m\right)\right]\nonumber
\end{eqnarray}
where $a$ is the mean square distance between neighboring beads
along the chain. It has been used that the force between the beads
can be obtained from the effective potential
$u=(3k_{B}T/2a^{2})\sum_{n=2}^{N}(\overrightarrow{x}_{n}-
\overrightarrow{x}_{n-1})^{2}$ which follows from the equilibrium
(Gaussian) distribution of the beads~\cite{3,4}. Due to the
dependence of the Oseen tensor on the difference
$\overrightarrow{r}_{nm}=\overrightarrow{x}(n)-\overrightarrow{x}(m)$,
Eq. (9) is nonlinear and thus hardly solvable analytically. We use
the common approximation of preaveraging of the tensor over the
equilibrium distribution $P(r_{nm})=(2\pi
a^{2}|n-m|/3)^{-3/2}\exp[-3r_{nm}^{2}/(2a^{2}|n-m|)]$:
\begin{equation}\label{eq10}
\left\langle H_{\alpha \beta nm}^{\omega} \right\rangle_{0}=
\delta_{\alpha\beta}h^{\omega}\left(n-m\right),
\end{equation}\begin{eqnarray}
h^{\omega}\left(n-m\right)=\frac{1}{\sqrt{6\pi^{3}|n-m|} \eta a
}\left[1-\sqrt{\pi}z e^{z^{2}}\text{erfc}(z)\right]\nonumber,
\end{eqnarray}
with $z\equiv\chi a \left(|n-m|/6\right)^{1/2}$. In the case
without memory [4] the function $h$ at large $|n-m|$ behaves as
$\sim|n-m|^{-1/2}$; now the effective interaction between the
beads disappears more rapidly, $\sim|n-m|^{-3/2}$. Since Eq. (9)
now contains only the diagonal terms, it can be solved using the
FT in the variable $n$,
$\overrightarrow{x}^{\omega}\left(n\right)=\overrightarrow{y}_{0}^{\omega}+2\sum_{p\geq
1}\overrightarrow{y}_{p}^{\omega}\cos(\pi np/N)$, where the
boundary conditions at the ends of the chain have been taken into
account~\cite{4}, $\partial \overrightarrow{x}/\partial n=0$ at
$n=0, N$. The inverse FT then yields the following equation for
the Fourier components
$\overrightarrow{y}_{p}^{\omega}$:\begin{equation}\label{eq11}
\overrightarrow{y}_{p}^{\omega}=\overrightarrow{f}_{p}^{\omega}\left[-i\omega
\Xi_{p}^{\omega}-M\omega^{2}+K_{p}\right]^{-1},
\end{equation}
where
$\Xi_{p}^{\omega}\equiv\xi^{\omega}\left[1+\left(2-\delta_{p0}\right)Nh_{pp}^{\omega}\right]^{-1}$,
and $K_{p}\equiv 3\pi^{2}p^{2}k_{B}T/(Na)^{2}$, $p=0,1,2,...$ The
matrix $h_{pp}^{\omega}$ is defined by the integral
\begin{equation}\label{eq12}
h_{pp}^{\omega}=\frac{1}{N^{2}}\int_{0}^{N}dn\int_{0}^{N}dm
h^{\omega}(n-m)\cos\frac{\pi pn}{N}\cos\frac{\pi pm}{N}.
\end{equation}
In obtaining Eq. (11) the fact  that the nondiagonal elements of
the matrix are small in comparison with the diagonal ones and can
be in the first approximation neglected has been already taken
into account; the substantiation of this is the same as in Refs.
[3,4]. Equation (11) can be investigated as it is usually done in
the theory of Brownian motion. One can use the
fluctuation-dissipation theorem (FDT)~\cite{18} or the properties
of the random forces~\cite{12}. The forces acting on different
beads $n$ and $m$ are uncorrelated, thus their correlator is
$\sim\delta_{nm}$. In going to the continuum approximation the
Kronneker symbol has to be replaced by the $\delta$-function,
$\delta(n-m)$, so in the FT we have
\begin{equation}\label{eq13}
\left\langle
f_{p\alpha}^{\omega}f_{q\alpha}^{\omega'}\right\rangle=
\frac{k_{B}T\text{Re}\Xi_{p}^{\omega}}{\left(2-\delta_{p0}\right)\pi
N}\delta_{\alpha\beta}\delta_{pq}\delta(\omega+\omega').
\end{equation}
Equation (11) then yields the following expression for the time
correlation function of the Fourier components $y_{\alpha p}$,
$\psi_{p}(t)=\left\langle y_{\alpha p}(0)y_{\alpha
p}(t)\right\rangle$:
\begin{equation}\label{eq14}
\psi_{p}(t)=\frac{k_{B}T}{\left(2-\delta_{p0}\right)\pi
N}\int_{-\infty}^{\infty}d\omega\frac{e^{-i\omega t
}\text{Re}\Xi_{p}^{\omega}}{\left|-i\omega\Xi_{p}^{\omega}-M\omega^{2}+K_{p}\right|^{2}},
\end{equation}
in agreement with the FDT. The generalized susceptibility is
$\alpha_{p}(\omega)=[(2-\delta_{p0})N]^{-1}[-i\omega\Xi_{p}^{\omega}-M\omega^{2}+K_{p}]^{-1}$,
and the generalized forces corresponding to the coordinates
$y_{p\alpha}^{\omega}$ are $Nf_{p\alpha}^{\omega}$. Using the
Kramers-Kronig relation~\cite{19}, the same initial value for the
function $\psi_{p}$ at $t=0$ as in the RZ theory is obtained:
$\psi_{p}(0)=k_{B}T\alpha_{p}(0)=k_{B}T(2NK_{p})^{-1}$, $p>0$.

Equation (14) gives the solution of the model. Knowing
$\psi_{p}(t)$, other correlation functions of interest can be
found, e.g. the VAF, $\Phi_{p}(t)=\left\langle v_{\alpha
p}(0)v_{\alpha p}(t)\right\rangle=-d^{2}\psi_{p}(t)/dt^{2}$, or
the mean square displacement (MSD), $\left\langle\Delta
y_{p}^{2}(t)\right\rangle=2[\psi_{p}(0)-\psi_{p}(t)]$. The
previous RZ results in the absence of memory are obtained by
putting $\omega=0$ in $\Xi_{p}^{\omega}$, Eq. (11), and $M=0$. The
mode $y_{0}$ describes the motion of the center of inertia of the
coil~\cite{3,4}. In the RZ case we get
$\psi_{0}(0)-\psi_{0}(t)=D_{C}t$. The diffusion coefficient
$D_{C}=k_{B}T\left(h_{00}^{0}+1/N\xi\right)$ contains the Zimm
($D_{C}=8k_{B}T\left(3\sqrt{6\pi^{3}N}\eta a\right)^{-1}$) and
Rouse ($D_{C}=k_{B}T/N\xi$) limits. The internal modes ($p>0$)
relaxed exponentially,
$\psi_{p}(t)=(k_{B}T/2NK_{p})\exp(-t/\tau_{p})$, with the
relaxation times
$\tau_{p}=\xi/\left[K_{p}\left(1+2Nh_{pp}^{0}\xi\right)\right]$,
where $h_{pp}^{0}=\left(12\pi^{3}Np\right)^{-1/2}(\eta a)^{-1}$.

\subsection{The Rouse limit with memory} Now let us consider the case with memory,
the limit of Rouse~\cite{15}. It assumes that the hydrodynamic
interaction contribution to $\Xi_{p}^{\omega}$ in Eq. (11) is
negligible for all $\omega$s. The subsequent equations then change
only by the substitution $\Xi_{p}^{\omega}\approx \xi^{\omega}$.
The corresponding integral in Eq. (14) is encountered in the
theory of Brownian motion of one particle: in the case when $p=0$,
the particle is free, and if $p>0$, it is in a harmonic field with
the force constant $K_{p}$. Such problems were solved in a number
of investigations beginning from the work~\cite{16}. Adopting the
known solutions, see e.g. Ref. [12], for the MSD of the coil we
have the following asymptotic
expression:\begin{equation}\label{eq15} \left\langle\Delta
y_{0}^{2}\left(t\right)\right\rangle=2D_{C}t\left[1-\frac{2}{\sqrt{\pi}}\sqrt{\frac{\tau_{b}}{t}}
+\frac{2}{9}\left(4-\frac{M}{M_{s}}\right)\frac{\tau_{b}}{t}-...\right],
\end{equation}
where $t>>\tau_{b}=b^{2}\rho/\eta$ and $M_{s}$ is the mass of the
solvent displaced by one bead. For small times $\left\langle\Delta
y_{0}^{2}\left(t\right)\right\rangle\approx
k_{B}Tt^{2}/N\left(M+M_{s}/2\right)$. (The physically correct
value $k_{B}Tt^{2}/NM$ can be obtained only if the compressibility
is taken into account~\cite{12,16}.) It is well seen from Eq. (15)
and confirmed by numerical calculations of Eq. (14) how slowly the
previous result is approached: the second ($\sqrt{t}$) term is
less than $1$ per cent of the first only for $t>10^{4}\tau_{b}$.
For long chains when the terms $\sim M$ can be neglected, at $t=10
\tau_{b}$ the MSD constitutes only about $3/4$ of the Einstein
limit. Note that for the motion of the coil as a whole the exact
analytical solution exists which differs from the known solution
for one Brownian particle only by a factor $1/N$ ~\cite{16,20}.
For example, the VAF of the center of mass of the Rouse coil is
\begin{equation}\label{eq16}
\Phi_{0}(t)=\frac{\Phi_{0}(0)}{\lambda_{1}-\lambda_{2}}
\sum_{i=1,2}(-1)^{i+1}\lambda_{i}e^{(\lambda_{i}^{2}t)}\text{erfc}(\lambda_{i}\sqrt{t}),
\end{equation}
where $\Phi_{0}(0)=k_{B}T/(M+M_{s}/2)N$ and $\lambda_{i}$ are the
(complex) roots of the equation
$\lambda^{2}+\sqrt{\tau_{b}}\lambda/\tau+1/\tau=0$ with
$\tau=(M+M_{s}/2)/\xi$.

The long-time asymptote of the function $\psi_{p>0}(t)$ describing
the relaxation of the Rouse modes is
\begin{equation}\label{eq17}
\frac{\psi_{p}(t)}{\psi_{p}(0)}=-\frac{1}{2\sqrt{\pi}}
\frac{\tau_{p}}{\tau_{b}}\left[\left(\frac{\tau_{b}}{t}\right)^{3/2}
+3\frac{\tau_{p}}{\tau_{b}}\left(\frac{\tau_{b}}{t}\right)^{5/2}+...\right].
\end{equation}

\subsection{The Zimm case}
For the Zimm model, when the hydrodynamic
interaction is strong for all frequencies that significantly
contribute to the studied correlation functions, we have in Eq.
(11) $\Xi_{p}^{\omega}\approx
\left[\left(2-\delta_{p0}\right)Nh_{pp}^{\omega}\right]^{-1}$. The
Oseen matrix (12) can be calculated with any degree of precision,
e.g. for $p=0$ we have the exact result
\begin{equation}\label{eq18}
h_{00}^{\omega}=\frac{3\sqrt{\pi}}{4\widetilde{z}}h_{00}^{0}
\left[1-\frac{2}{\sqrt{\pi}\widetilde{z}}-\frac{1}{\widetilde{z}}
\left(e^{\widetilde{z}^{2}}\text{erfc}\widetilde{z}-1\right)\right],
\end{equation}
where $\widetilde{z}\equiv(N\chi a/6)^{1/2}$. Then in the
$t$-representation the correlation function can be obtained using
the standard methods~\cite{12}. Here we show the main terms of the
asymptotic expansion of the time correlation functions that can be
found also by using the expansion of the susceptibility
$\alpha_{p}(\omega)$ in small $(-i\omega)^{1/2}$. In the case of
the diffusion of the coil as a whole we obtain the expression
similar to Eq. (15),
\begin{equation}\label{eq19}
\psi_{0}(0)-\psi_{0}(t)=D_{C}\left[t-\frac{2}{\sqrt{\pi}}\left(\tau_{R}t\right)^{1/2}+...\right].
\end{equation}
The characteristic time in this equation is
$\tau_{R}=R^{2}\rho/\eta$. The form of Eq. (19) exactly
corresponds to that familiar in the theory of Brownian motion of
rigid particles of radius $R$. Here $R$ is the hydrodynamic radius
of the coil~\cite{3,4}, determined from the relation
$D_{C}=k_{B}T/(6\pi R\eta)$ where $D_{C}$ is the Zimm diffusion
coefficient. In the standard model of Gaussian chains the radius
$R$ is given also by the Kirkwood relation~\cite{3}
$R^{-1}=N^{-2}\sum_{n=1}^{N}\sum_{m=1,m\neq n}^{N}\langle
r_{nm}^{-1}\rangle \approx 8\sqrt{2/3\pi N}/a$, where $r_{nm}$ is
the distance between the beads and $N>>1$.

Consider now the internal modes of the polymer, $p>0$. The
components of the matrix (12) are calculated as in Ref. [4] and
can be expressed through special functions, so that their
expansion is known to any desired power of $\sqrt{-i\omega}$. The
first correction to the classical result is determined by the
coefficient at the term $~(-i\omega)$. The first nonvanishing
correction to the susceptibility $\alpha_{p}(\omega)$ is given by
the term $~(-i\omega)^{5/2}$. The coefficient at this term is
determined by the expansion coefficients at the terms $(-i\omega)$
and $(-i\omega)^{3/2}$ in the expansion of $h_{pp}^{\omega}$. For
the correlation function $\psi_{p>0}(t)$, using Eq. (14), we thus
find
\begin{equation}\label{eq20}
\frac{\psi_{p}(t)}{\psi_{p}(0)}\approx
-\frac{2^{9}}{45\pi^{3}}\sqrt{\frac{2}{\pi}}\left(1
+\frac{16}{3\pi^{2}p}\frac{\tau_{R}}{\tau_{p}}\right)\frac{1}{p^{3}}
\frac{\tau_{p}\tau_{R}^{3/2}}{t^{5/2}},
\end{equation}
where $t>>\tau_{R}$ and $\tau_{p}=(Na^{2})^{3/2}(\eta/k_{B}T)(3\pi
p^{3})^{-1/2}$ is the Zimm relaxation time~\cite{3,4}.

\subsection{End-to-end vector and the dynamic structure factor}
Having the solutions (17) and (20), the
evolution of the end-to-end vector of the chain can be
investigated. From the relation
$\overrightarrow{R}(t)=\overrightarrow{x}(t,N)-\overrightarrow{x}(t,0)=-4\sum_{p=1,3,...}
\overrightarrow{y}_{p}(t)$ one finds
\begin{equation}\label{eq21}
\Phi(t)=\left\langle
\overrightarrow{R}(t)\overrightarrow{R}(0)\right\rangle=48\psi_{1}(0)\sum_{p=1,3,...}\frac{1}{p^{2}}\frac{\psi_{p}(t)}{\psi_{p}(0)}
\end{equation}
for both the models. For the Rouse model at long times
\begin{equation}\label{eq22}
\Phi(t)=-\frac{\pi^{4}}{4\sqrt{\pi}}\frac{\tau_{1}\tau_{b}^{1/2}}{t^{3/2}}
\psi_{1}(0)\left(1+\frac{3\pi^{2}}{10}\frac{\tau_{1}}{t}+...\right),
\end{equation}
and in the Zimm case the long-time asymptote reads
\begin{equation}\label{eq23}
\Phi(t)\approx
-3\sqrt{2\pi}\psi_{1}(0)(\tau_{R}+1.85\tau_{1})\tau_{R}^{3/2}t^{-5/2}.
\end{equation}
(In Eqs. (22) and (23) $\tau_{p=1}$ is the relaxation time for the
corresponding model.)

Finally, we give the result for the intermediate scattering
function $G(\overrightarrow{k}t)=N^{-1}\sum_{nm}\langle
\exp\{i\overrightarrow{k}[\overrightarrow{x}_{n}(t)-\overrightarrow{x}_{m}(0)]\}\rangle$
that is used in the description of the dynamic light or neutron
scattering from a polymer coil~\cite{3} ($\overrightarrow{k}$ is
the wave-vector change at the scattering). Acting in a similar way
as in Ref. [3], $G(\overrightarrow{k},t)$ can be for large $t$
approximated by the expression
\begin{eqnarray}\label{eq24}
G(k,t)\approx N
\exp\left\{-k^{2}\left[\psi_{0}(0)-\psi_{0}(t)\right]\right\}\nonumber\\
\times\exp\left[\frac{-Na^{2}k^{2}}{36}\left(1-\frac{8N^{2}a^{4}k^{4}}{3\pi^{6}}
\sum_{p=2,4,...}^{\infty}\frac{1}{p^{6}}\frac{\psi_{p}(t)}{\psi_{p}(0)}
\right)\right]
\end{eqnarray}
valid for $kR<<1$ (in the opposite case, the function $G(k,t)$
becomes very small at long times). Equation (24) is equally
applicable for both considered models, if the corresponding Eqs.
(15) and (17) for the Rouse case, or (19) and (20) for the Zimm
model are substituted here. One can see that the contribution of
the internal modes is small and thus hardly detectable against the
diffusion term given by the first exponential. However, our
predictions concerning the diffusion of the coil as a whole,
\begin{equation}\label{eq25} G(k,t)\approx N
\exp\left\{-k^{2}\left[\psi_{0}(0)-\psi_{0}(t)\right]\right\},
\end{equation}
could be directly measured in the scattering experiments.

\section{Conclusion}
We conclude that in the generalized RZ model, when the memory of
the viscous solvent is taken into account, the relaxation of the
correlation functions describing the polymer motion essentially
differs from the original theory. The MSD is at short times $\sim
t^{2}$ (instead of $\sim t$). At long times it contains additional
(to the Einstein term) contributions, the leading of which is
$\sim\sqrt{t}$. The internal normal modes of the polymer motion
now do not relax exponentially. It is well displayed in long-time
tails of their time correlation functions. The longest-lived
contribution to the correlation function of the bead displacement
is $\sim t^{-3/2}$ in the Rouse limit and $t^{-5/2}$ in the Zimm
case, when the hydrodynamic interaction is strong. It would be
interesting to investigate the found peculiarities using computer
simulation methods and experimentally, e.g. by the dynamic light
or neutron scattering. Simple estimations show that the relaxation
of the internal modes, $\psi_{p>0}(t)$, although qualitatively
different from the previous RZ model, for real polymers only
slightly differs from the traditional exponential law
$\sim\exp(-t/\tau_{p})$, except at high frequencies (it is because
the relaxation time for the internal modes, $\tau_{p}$, is much
larger than the characteristic times $\tau_{b}$ and $\tau_{R}$).
However, the differences from the original model, at least for the
Zimm case, should be experimentally accessible. Due to the
long-range character of the hydrodynamic field, the characteristic
time of the Zimm model, $\tau_{R}$, is determined by the size of
the whole coil. For a typical radius around $100$ nm, the density
and viscosity of water at room conditions, one gets $\tau_{R}$
about $10$ nanoseconds ($\tau_{R}/\tau_{p=1}\approx5\times10^{-4}$
sec). Taking into account the possibilities of current experiments
(e.g. in Ref.~\cite{21} the dynamic structure factor of polymers
in solution was studied using quasi-elastic light scattering in a
time window beginning from 12.5 ns, and even shorter times are
accessible by the neutron spin-echo technique~\cite{22}), and the
fact that the function $\psi_{0}(0)-\psi_{0}(t)$ approaches the
Einstein limit $D_{C}t$ very slowly as $t$ increases, the
nondiffusive ("ballistic") motion of the polymer coil should be
readily observable by the dynamic light and neutron scattering. In
fact, similar experiments were successfully carried out on single
Brownian particles. For example, using the diffusive wave
spectroscopy the ballistic motion of polystyrene spheres with the
radius $b=0.206 \mu$m in aqueous solution (with the characteristic
time $\tau_{R}$ about $0.04 \mu$sec) was observed~\cite{23}. The
size of such particles corresponds to the hydrodynamic radius of
the DNA coil of a molecular weight $6\times10^{6}$ g/mol (with the
diffusion coefficient $D_{C}\approx1.3\times10^{-8}$
 cm$^{2}$/s)~\cite{24}. The nondiffusive motion of even smaller
(with a radius $<100$ nm) particles was observed in the
experiments~\cite {25}. It has been found in these and other works
that the apparent diffusion coefficient of Brownian particles is
smaller than that following from the Einstein theory. These
observations are very similar to the situation described by us for
the polymer coils. The tails in the MSD lead to a slower decay of
the dynamic structure factor (25). This corresponds to diffusion
with an effectively smaller diffusion coefficient (as well as the
first cumulant) than predicted by the previous theory. This is the
long-standing unresolved "puzzle" between the theory and
experiments. We believe that the proposed theory could help to
solve this problem in the description of the dynamic scattering
experiments on polymers~\cite{5,6,7} and thus to contribute to a
deeper understanding of the dynamical properties of polymers.

\section{Acknowledgment} AVZ and VL thank the NWO (Dutch Research
Council) for grants that enabled them to visit the Leiden
Institute of Chemistry where a part of this work was done. We are
grateful to Prof. D. Bedeaux and Dr. A.V. Zvelindovsky for kindest
hospitality and fruitful discussions. This work was supported by
the grant VEGA, Slovak Republic. \\\\
$^{a)}$Author to whom correspondence should be addressed.
Electronic mail: lisy@upjs.sk

\end{document}